\documentclass[review]{elsarticle}

\usepackage{lineno,hyperref}
\modulolinenumbers[5]

\journal{Journal of \LaTeX\ Templates}

\graphicspath{{figures/}{pictures/}{images/}{./}} 

\usepackage{color,soul}

\begin{document}

\begin{frontmatter}

\title{Exploring the Design Space of Immersive Urban Analytics}

\author[hkust]{Chen Zhu-Tian}
\ead{zhutian.chen@outlook.com}

\author[zju]{Yifang Wang}
\author[zju]{Tianchen Sun}
\author[hnu]{Xiang Gao}
\author[zju]{Wei Chen}
\author[hnu]{Zhigeng Pan}
\author[hkust]{Huamin Qu}

\author[zju]{Yingcai Wu~\corref{mycorrespondingauthor}}
\cortext[mycorrespondingauthor]{Corresponding author}

\address[hkust]{Hong Kong University of Science and Technology}
\address[zju]{Zhe Jiang University}
\address[hnu]{Hangzhou Normal University}

\begin{abstract}
    Recent years have witnessed the rapid development and wide adoption of immersive head-mounted devices,
    such as HTC VIVE, Oculus Rift, and Microsoft HoloLens. 
    These immersive devices 
    have the potential to significantly
    extend the methodology of urban visual analytics by providing critical 3D context information
    and creating a sense of presence. 
    In this paper, we propose an theoretical model to characterize the visualizations in immersive urban analytics.
    Furthermore, based on our comprehensive and concise model, 
    we contribute a typology of combination methods of 2D and 3D visualizations 
    that distinguish between \emph{linked views, embedded views}, and \emph{mixed views}. 
    We also propose a supporting guideline to assist users in selecting a proper view under certain 
    circumstances by considering \emph{visual geometry} and \emph{spatial distribution} of the 2D and 3D visualizations.
    Finally, based on existing works, possible future research opportunities are explored and discussed.
\end{abstract}

\begin{keyword}
Immersive Urban Analytics, Virtual/Augmented/Mixed Reality, Urban Visualizations, Information Visualizations, Visualization Theory
\end{keyword}

\end{frontmatter}



\section{Introduction}
\label{section:introduction}

Urban visual analytics has been proven useful in solving various problems of urban cities, 
such as location selection~\cite{urban-vis-smartadp}, urban planning~\cite{urban-line-trajgraph}, and traffic analysis~\cite{urban-guosiming}, 
by integrating the computational power of machines and the domain knowledge of experts.
In urban visual analytics, visual representations of urban data provide a crucial
context for exploration and analysis~\cite{geovisual-analytics-for-spatial-decision-making}.

Most existing studies of urban visual analytics utilize 2D maps~\cite{wenchao-survey} on which every 
point is viewed overhead. 
As 2D maps create an abstraction of the real world, 
the maps lose significant context information on the urban environment, 
consequently leading to
the severe limitations in solving space-related problems in the urban context.
First, the lack of depth information of vertical cities poses a significant challenge for making informed
decisions in many scenarios. For example, selecting befitting locations to place billboards
exclusively based on traffic flow on 2D maps would be difficult for advertising managers~\cite{urban-vis-smartadp},
because candidate locations in vertical cities may be near or under buildings, overpasses, utility wires, and so on.
Second, a 2D map that lacks the appearance of the real world cannot provide users with a sense of presence.
For example, in a 2D map, both magnificent skyscrapers and tiny bungalows are displayed as polygons.
In such a lack of a sense of presence, users cannot fully apply their expertise and domain knowledge
to making a confident spatial decision. Expensive field studies are frequently employed.
Therefore, growing interest has been observed in applying 3D maps for urban
visual analytics~\cite{urban-3d-urbane, urban-3d-visaware}.

In recent years, various of immersive head-mounted devices, such as HTC VIVE, Oculus
Rift, and Microsoft HoloLens, have been invented and adopted in a wide range of settings.
The immersive devices use stereoscopic techniques to create a natural support for 3D display, 
thereby creating an engaging and immersive visual environment~\cite{ia-position-paper}.
The significant development and broad adoption of the immersive devices shed new light on visualizing
heterogeneous geospatial urban data in such an immersive environment; this field can be referred to  as 
\emph{immersive urban analytics}.

Recently, researchers from urban visual analytics~\cite{space-time-visual-anlytics}
and immersive analytics~\cite{ia-position-paper} have raised 
%
questions on how to visualize the abstract data together with 3D models. 
Abstract data are commonly visualized in a 2D manner, since the 3D display of the data remains controversial and may cause ambiguity~\cite[Chapter~6]{tamara-visualization}.
By contrast, as  a type of physical data, city models can be naturally displayed in 3D. 
It remains unclear that how to seamlessly display 2D abstract data together with 3D physical city models in an effective way.

To address this issue, we first summarize an abstract model to characterize the visualization in immersive analytics.
Based on our model and the fundamental theorem of Euclidean geometry, 
we propose an innovative typology classifying the ways to visually integrate physical
and abstract data into three categories, namely, \emph{Linked view, Embedded view}, and \emph{Mixed view}.
Furthermore, we conduct preliminary explorations and summarize two plain and comprehensive design considerations, namely, \emph{Visual geometry} and \emph{Spatial distribution}, to assist designers in effortlessly choosing the best view under certain circumstances.
We demonstrate the effectiveness of our design considerations with several examples.

\section{Terminology}
\label{section:relatedwork}
Immersive technologies have been researched for several decades.
We only focus on the emerging immersive technologies.
These technologies can create various immersive environments which are different from each other.
In this section,
we first introduce the characteristics of  different immersive environments,
then we summarize the types of urban data which will be visualized in immersive environments.

\subsection{Immersive Environments}
Immersive environments are the environments created by immersive technologies~\cite{ia-position-paper}.
Immersive environments can be roughly classified into three categories, namely, Virtual Reality (VR), Augmented Reality (AR), and Mixed Reality (MR).
Several existing works argued and gave formal definitions of VR, AR, and MR.
However, since the immersive industry grows rapidly in recent years,
most of these works are obsolete.
Rather than re-define the concept of these environments,
we prefer to identify their distinguishable characteristics,
which can help us better understand the environments for visualization.

 \begin{figure}[tb]
 	\centering 
 	\includegraphics[width=\columnwidth]{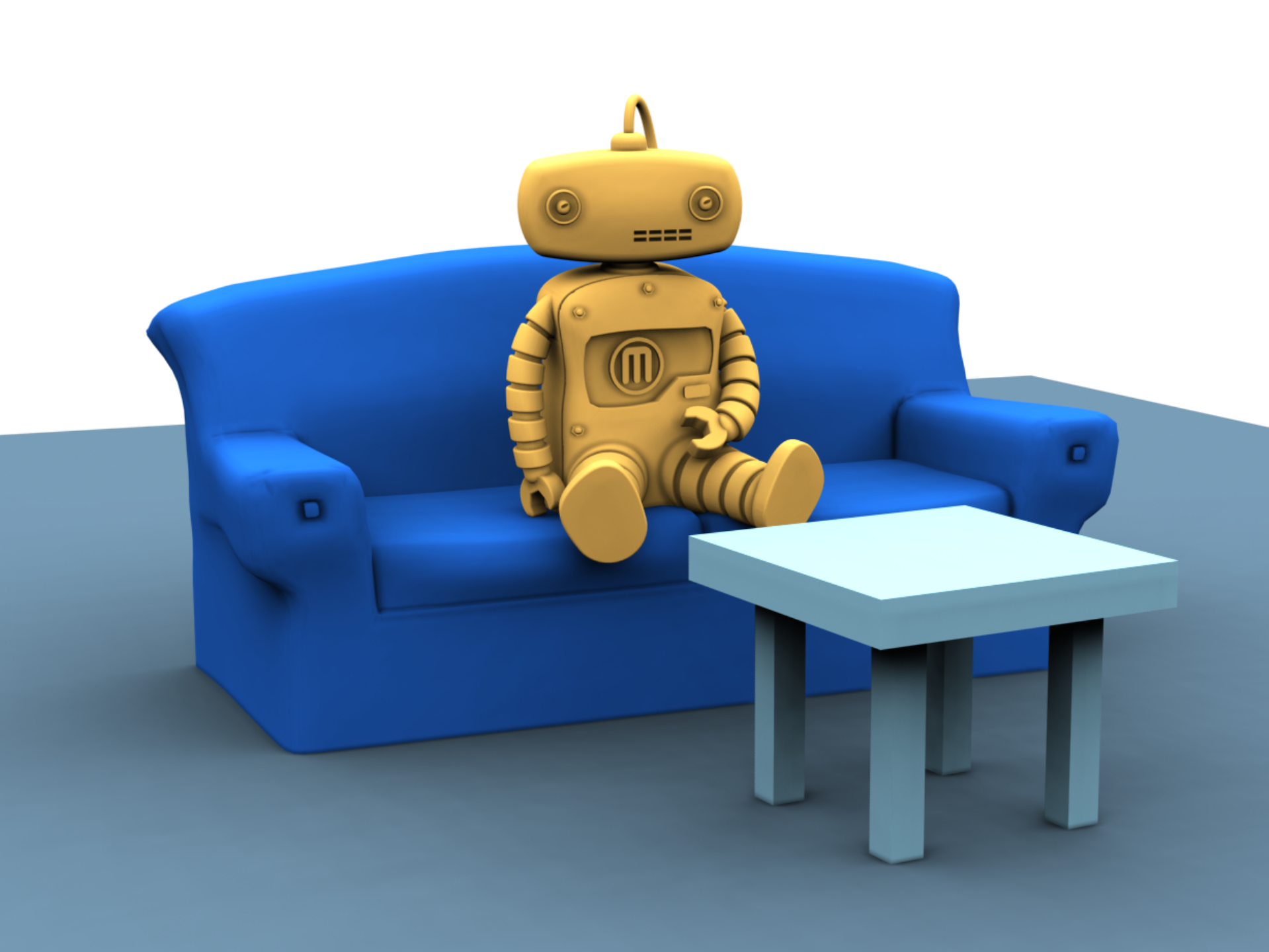}
 	\caption{The virtual reality environments entirely immerse the users in a virtual space, blocking the real world surroundings.}
 	\vspace{-5mm}
 	\label{fig:vr}
 \end{figure}
 
\textbf{Virtual Reality} We specify the virtual reality (VR) environments as environments
created by VR head-mounted display devices, 
such as HTC Vive, Oculus Rift, Samsung Gear, and Google Cardboard. 
The cave automatic virtual environments (CAVE) system, 
which can also create VR environments, 
is exclusive from our discussion because it is not an emerging technology.
The key characteristic of the VR environments is that they fully immerse the users in a digital world
and entirely occlude the natural surroundings.
Figure~\ref{fig:vr} is an example of VR environments.
Users can only see the virtual robot sits on a virtual sofa behind a virtual table.


 \begin{figure}[tb]
 	\centering 
 	\includegraphics[width=\columnwidth]{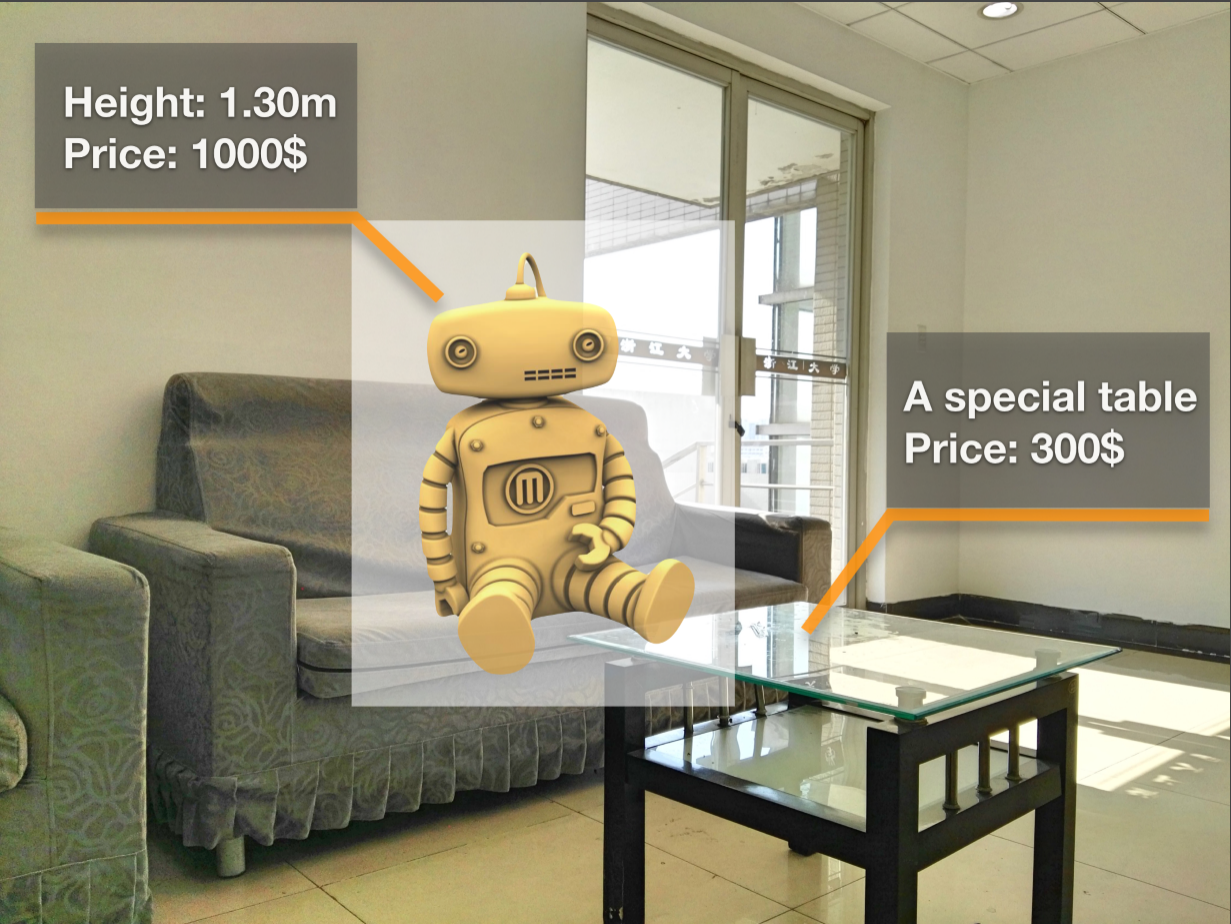}
 	\caption{The augmented environments directly overlay the digital content on top of the real world, allowing users to explore additional information.}
 	\vspace{-5mm}
 	\label{fig:ar}
 \end{figure}

\textbf{Augmented Reality} 
Azuma et al. defined AR~\cite{ar-survey} as systems that have the following three
characteristics: 1) combine the real and the virtual, 2) interactive in real time, and
3) (virtual content is) registered in 3D. This definition is well-accepted.
However, in recent year, many famous products (e.g. Pokeman Go, Google Glass),
which directly overlay the virtual content on top of the real world rather than register them in 3D,
claim themselves as AR.
The popularity of these products, especially the mobile and tablets applications,
makes consumers misunderstand the concept of AR,
thereby leading to a gap between academia~\cite{ar-survey,acadamia:avm1, acadamia:avm2, acadamia:avm3, acadamia:avm4} and industry~\cite{industry:vma1,industry:vma2, industry:vma3, industry:vma4, industry:vma5}.
In this work, we follow the trend of industry and specify the AR environments as the environments
in which the 2D or 3D virtual content is directly overlaid on top of the real world,
surrounding users with additional information.
Figure~\ref{fig:ar} shows an example of AR environments.
Users can see a virtual robot with additional digital information overlaied on the real world through the AR system.

 \begin{figure}[tb]
 	\centering 
 	\includegraphics[width=\columnwidth]{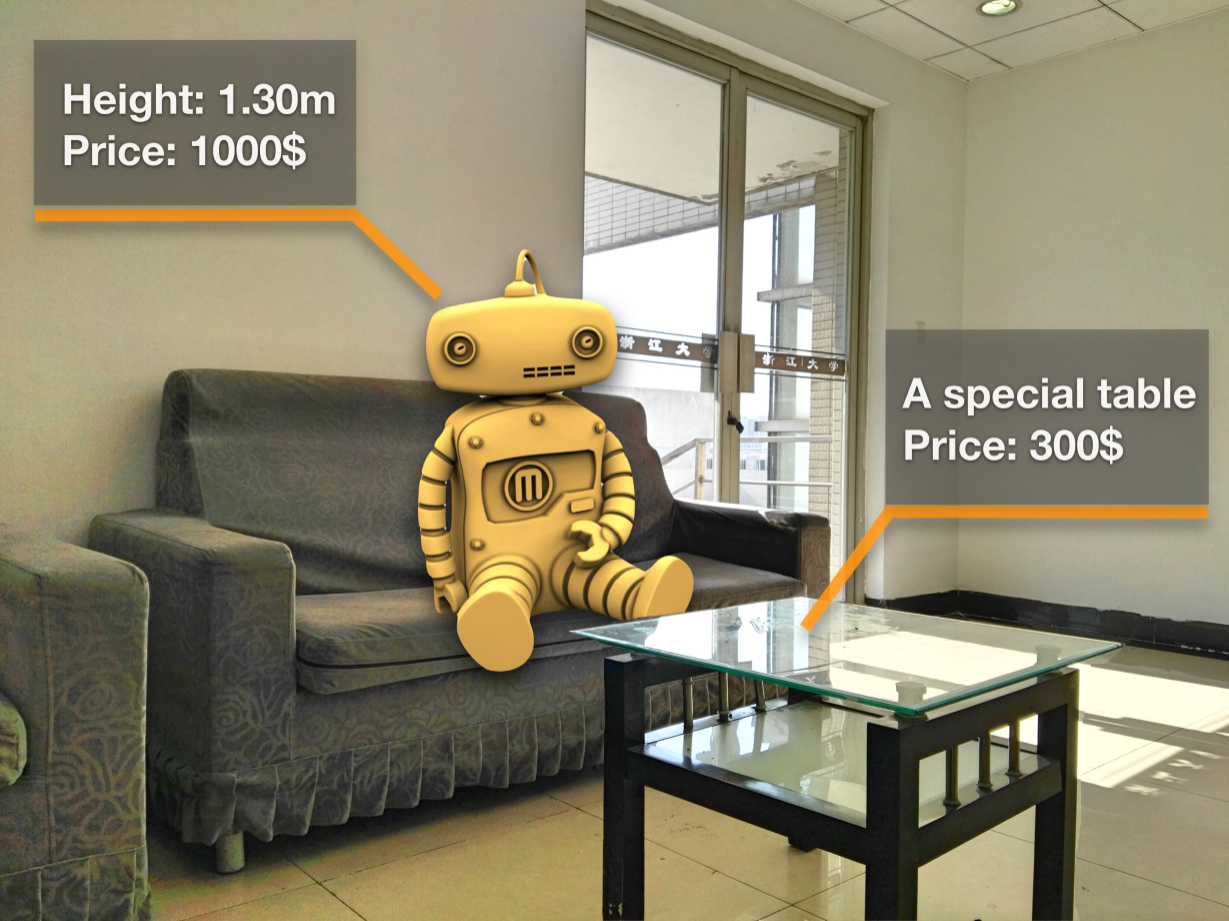}
 	\caption{The mixed environments blend the virtual content into the real world, enabling the virtual content responding to the real world.}
 	\vspace{-5mm}
 	\label{fig:mr}
 \end{figure}

\textbf{Mixed Reality Environment} To distinguish from the popular AR products,
some `real' AR products, such as HoloLens and Magic Leap, promote and define themselves using the term Mixed Reality (MR).
In the MR environments, the virtual objects are placed into the real world space realistically.
The virtual objects look, sound, and behave like real objects that are able to interact with the environment around them and also each other.
In a sense, the virtual objects in mixed reality are real objects without a tangible body.
From the perspective of technology, MR is harder than AR and VR.
Figure~\ref{fig:mr} demonstrates an example of MR environments.
Users can observe that the virtual robot is sitting on a real sofa.
His left leg is blocked by the table. Such kind of spatial occlusion makes it look like a real robot.
Additional digital information of the robot and real objects is also displayed.

\subsection{Urban Data}

In addition to the immersive environments, 
another key feature of immersive urban analytics is the content--that is to say the data to display.
In immersive urban analytics, the data of an object can be divided into two types,
namely, physical and abstract data.

\textbf{Physical Data} The physical data identifies and depicts an object in the physical world,
which answers questions such as \emph{``what does the building look like?''} and \emph{``where is the building?''}
This kind of data, which is related to 3D phenomena and visualized in a 3D manner,
is the primary research object in scientific visualization.
Using physical data, we can reconstruct the geometry and appearance of a building.

\textbf{Abstract Data} The abstract data describes the properties of an object, which answers questions 
such as \emph{``how much is the building?''} and \emph{``what is the population density of this region?''}
Abstract data, usually visualized in a 2D manner,
is the primary research object in information visualization and visual analytics.
Abstract data can help users  discover hidden knowledge.

The spatial representation is the major difference between physical and abstract data in which the former is given,
whereas the latter is selected~\cite{pitfalls-infovis}.

\section{Typology}
\label{section:design}

In immersive urban analytics, physical and abstract data are visualized: the map is created
based on physical data, whereas other spatial-temporal data are displayed for visual analytics.
Integrating the visualization of physical and abstract data has always been a critical topic in immersive analytics.
Additionally, the real objects, which can be interacted with the digital content,
may be part of the visualization, thereby leading to difficulties in creating holistic visualizations.
To address this problem, 
we first propose a model to characterize the visualization in immersive urban analytics.
Based on this model, we then deduce a typology that classifies the ways in which physical and abstract data are visually
integrated into three categories, namely, \textbf{linked view, embedded view}, and \textbf{mixed view}.
Finally, we propose two novel design considerations, specifically, \textbf{visual geometry} and \textbf{spatial distribution},
to guide a visualization developer to design a hybrid physical/abstract data visualization.

\subsection{Visualization Model of Immersive Urban Analytics}

To realize an effective way of visually integrating physical and abstract data,
we first require identifying the manners through which we can achieve the integration.
We attempt to consider this problem from an abstract aspect rather than using an exhaustive method,
by proposing a theoretical model to depict the visualization in immersive urban analytics.

\begin{figure}[tb]
	\centering 
	\includegraphics[width=\columnwidth]{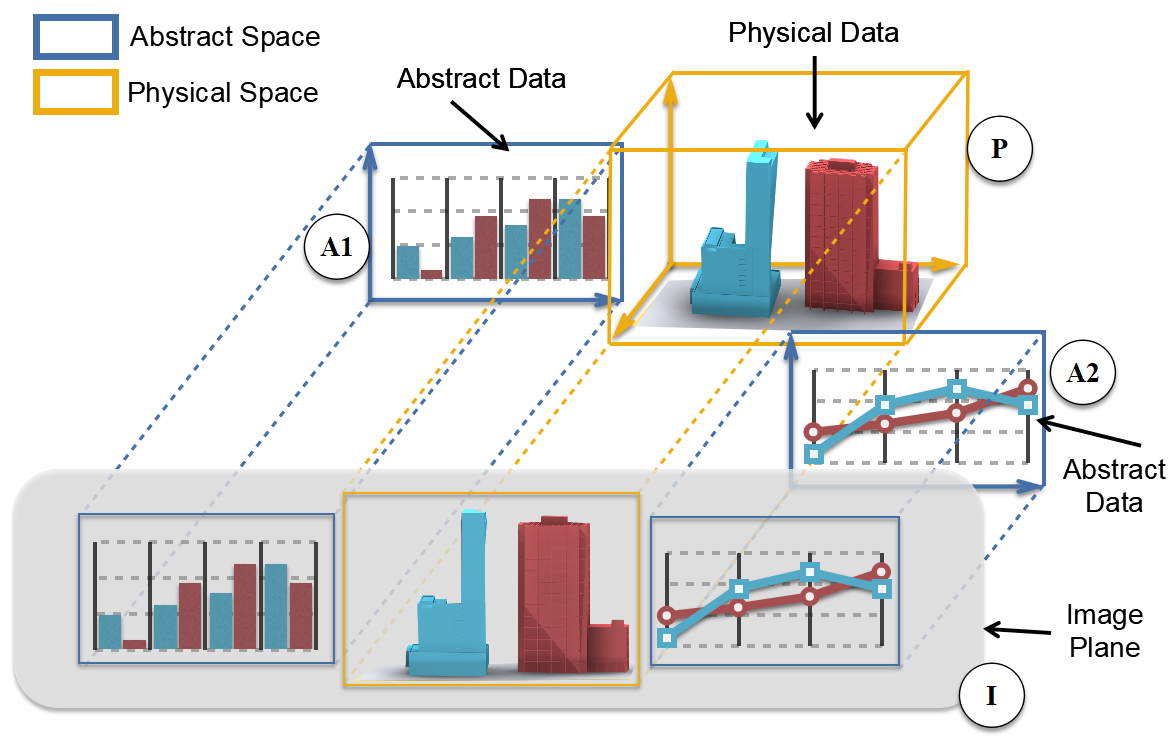}
	\caption{The model depicts the visualization in immersive urban analytics.
		Both \emph{A1} and \emph{A2} are 2D abstract spaces, in which the abstract data are rendered.
		\emph{P} is the 3D physical space, in which the physical data is rendered.
		Users perceive the data from the projections of spaces on the image plane \emph{I}.
	}
	\vspace{-5mm}
	\label{fig:model-vr}
\end{figure}

It is well known that physical data is usually visualized in 3D and abstract data is visualized in 2D.
In other words, there exist two kinds of rendering spaces in the visualization of immersive urban analytics,
namely, three-dimensional space and two-dimensional space.
We refer to the three-dimensional space as the \emph{physical space},
and the two-dimensional one as the \emph{abstract space}.
The definitions demonstrate the rendering spaces of data in immersive urban analytics.
However, it still remains unclear that how users perceive these data.
According to Ware~\cite[Chapter~5]{colin-visualization} and Munzner~\cite[Chapter~6]{tamara-visualization},
most of the visual information we perceive is on a 2D image plane, whereas the third
depth dimension is only a tiny additional fraction beyond the 2D image plane. 
Thus, no matter in which form the information is displayed in the spaces (2D or 3D), 
we perceive this information as a two-dimensional picture.
From the perspective of the human vision system,
only the projections on the image plane of the physical and abstract spaces matter.
Figure~\ref{fig:model-vr} illustrates our model: 
in an immersive urban analytics system,
the physical space (\autoref{fig:model-vr} P) is used for rendering physical data in 3D, whereas
the abstract spaces (\autoref{fig:model-vr} A1 and A2) are used for rendering abstract data in 2D.
The users perceive the visualization on an image plane (\autoref{fig:model-vr} I).

\begin{figure}[tb]
\centering 
\includegraphics[width=\columnwidth]{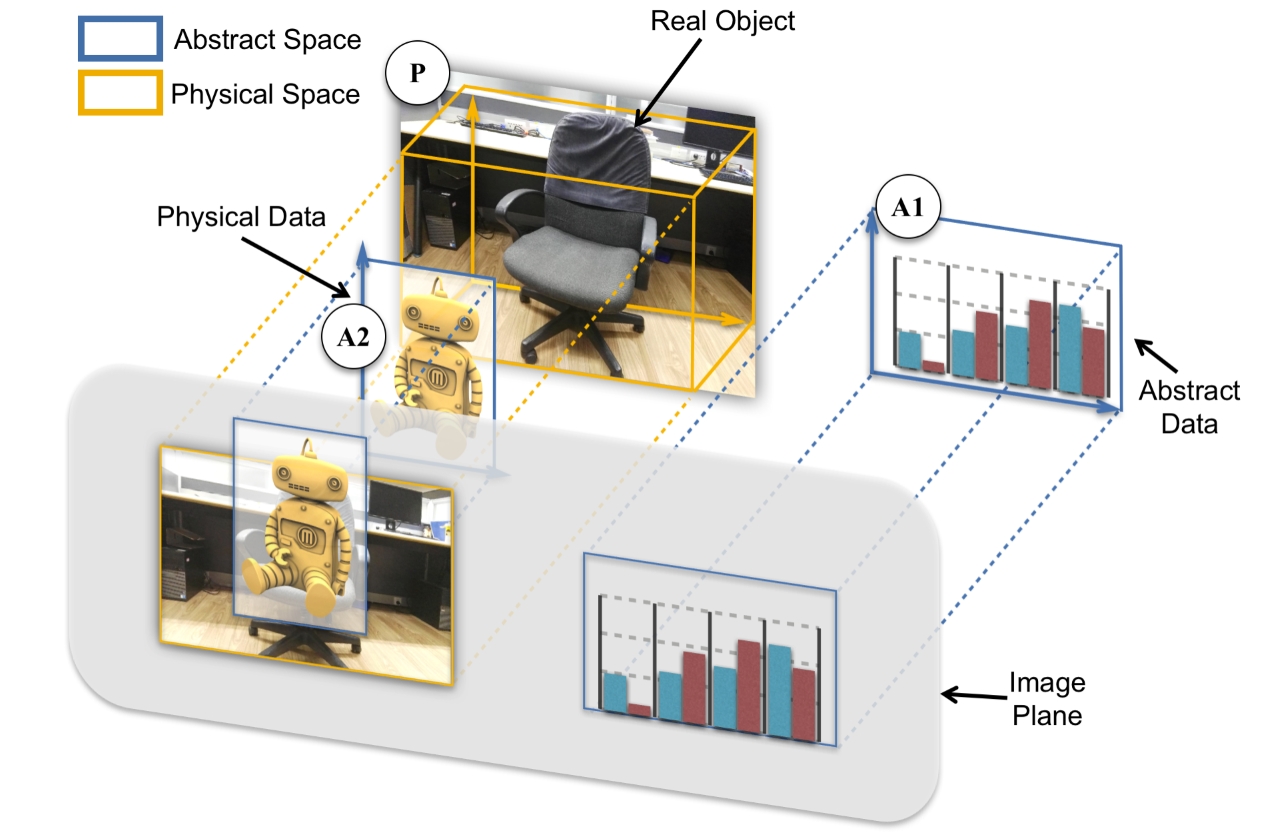}
\caption{The AR environments depicted by our model.
Both \emph{A1} and \emph{A2} are abstract spaces. \emph{P} is the physical space.
The virtual robot, which is the physical data, 
is rendered in \emph{A2} and directly overlaid on top of the \emph{P}.
}
\vspace{-5mm}
\label{fig:model-ar}
\end{figure}

\begin{figure*}[tb]
\centering 
\includegraphics[width=\columnwidth]{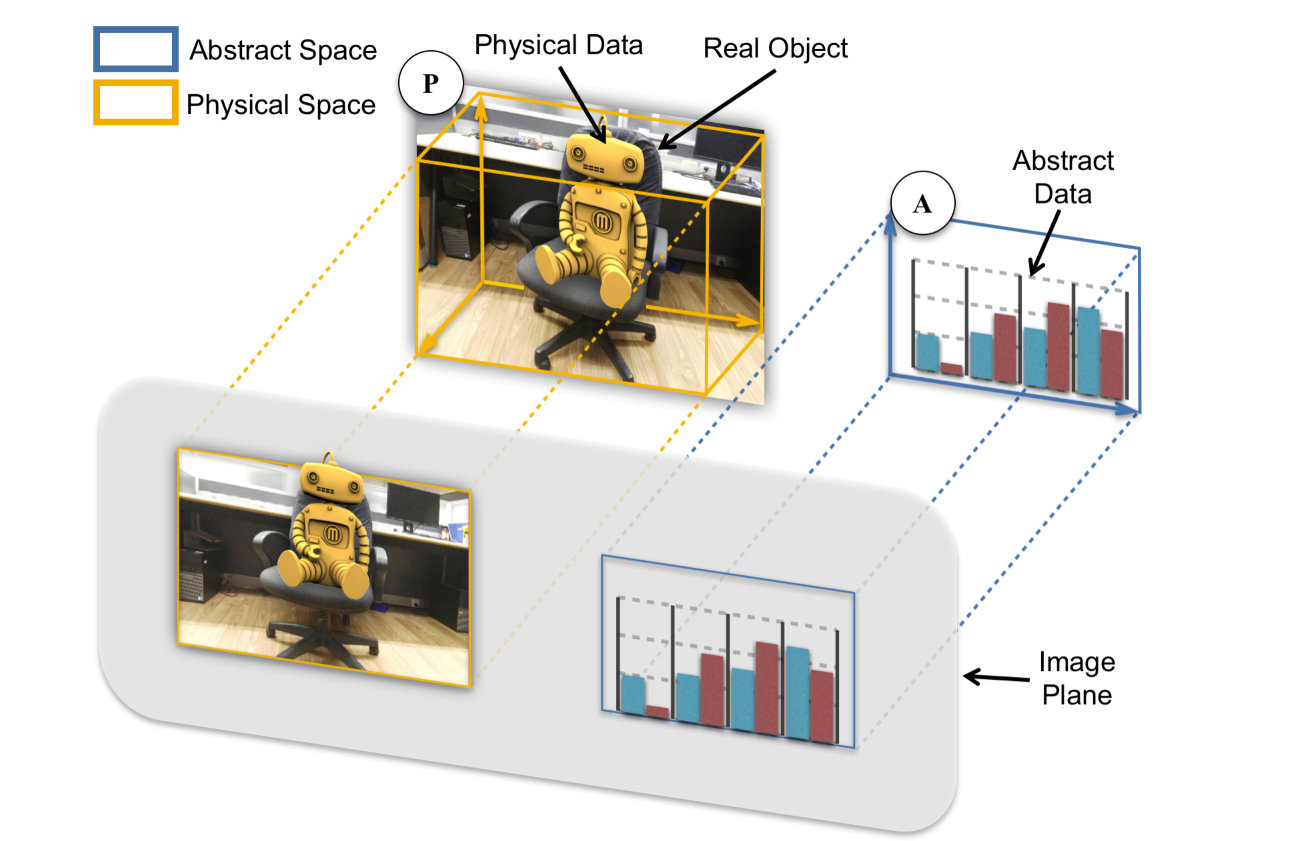}
\caption{The MR environments depicted by our model.
\emph{A} is a abstract space.
\emph{P} is the physical space, in which both the physical data and real objects are displayed.
}
\vspace{-5mm}
\label{fig:model-mr}
\end{figure*}

With this model, we can characterize VR, AR, and MR environments and depict the visualizations within them:

\begin{itemize}
    \item The VR environments is the basic condition (for our model) since there are no real objects exist.
    Only the physical and abstract data, which are displayed in physical and abstract spaces, 
    respectively, need to be considered.
    The VR environments is the default situation for our model. \autoref{fig:model-vr}
    presents an example of VR environment depicted by our model.
    In~\autoref{fig:model-vr}, both \emph{A1} and \emph{A2} are abstract spaces.
    They are in 2D spaces in which the abstract data is rendered.
    \emph{P} is the physical space, which is 3D space where the physical data is rendered.
    \item In the AR environments, the space where the real objects exist is a physical space
    since the real world is three dimensional.
    Although the physical data is usually displayed in 3D,
    in the AR environment, the virtual content is displayed in 2D and overlaid directly on top of the real objects,
    which means both the physical data and abstract data are displayed in the abstract space.
    For example, in \autoref{fig:model-ar},
    the virtual robot, which is the physical data, 
    is rendered in \emph{A2} and directly overlaid on top of the physical space \emph{P}.
    \item In the MR environments, the physical data is rendered and blended with the reality,
    and displayed in the real world, which is a physical space.
    Only the abstract data is rendered in the abstract space.
    An example is presented in~\autoref{fig:model-mr}. The virtual robot, which is a kind of physical data,
    and the real chair are displayed in the physical space \emph{P}. 
\end{itemize}
In a word, our model is comprehensive and can cover all environments of immersive urban analytics.
Next we will show how we can use this model to deduce a typology that classify the ways to 
visually integrate the physical and abstract data.

\subsection{Typology of Visual Integration of Physical and Abstract Data}

Based on our concise model, we transfer the problem of visually integrating the physical and the abstract data
to the problem of integrating the projections of the physical and abstract spaces.
Thus, we can sort out the basic methods from an illustrative detail of integration methods.

We assume that there are two spaces in the visualization to simplify our discussion.
Given that the size of the projection of a space is finite,
the two projections of the two spaces are considered to be two faces on an image plane. 
According to the fundamental theorem of Euclidean geometry, three relationships are observed between
two faces on a 2D plane, namely, separation, intersection, and adjacency.
Based on these three relationships, we classify the ways to visually integrate physical and abstract data as 
\emph{linked view, embedded view},and \emph{mixed view}.

We only consider the relationships from the perspective of design purpose. 
Specifically, the classification is based on the initial/default relationships of the two spaces' projections,
rather than the relationships in real time,
since in the immersive environments the users' viewpoints change frequently and
the relationships between the projections of spaces are changed accordingly.
Specifically, the projections between two spaces might separate, intersect, or juxtapose to each others based on the users' viewpoints.
We refer these relationships as to semantic relationships to distinguish from the spatial relationships.
Waqas et al. proposed a theoretical model~\cite{multi_coordinate}
which adopts a few similar concepts such as juxtaposition and superimposition.
However, their model aims to identify the design space of coordinated multiple views of abstract data
 in desktop environments based on the spatial relationships of views.
Hence their model is different from ours and it is not suitable for the immersive urban analytics.
We denote our work as a typology rather than a taxonomy because the former is appropriate for 
classifying abstract concepts, while the latter is appropriate for classifying empirically
observable events~\cite{taxonomy-vs-topology}.

\begin{figure*}[tb]
 \centering 
 \includegraphics[width=\columnwidth]{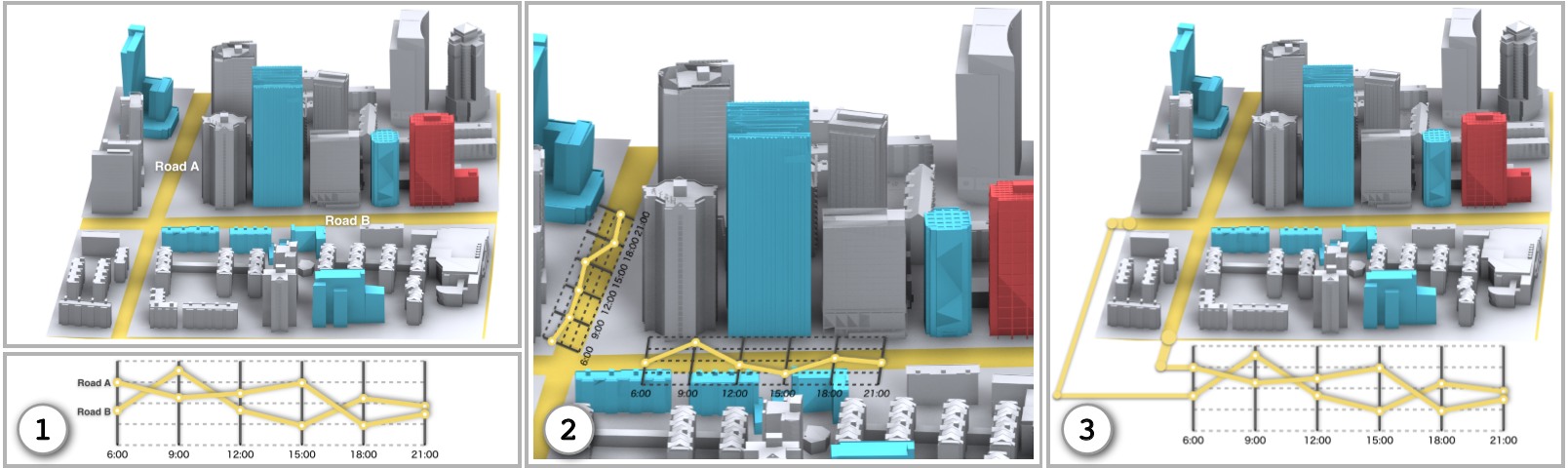}
 \caption{Our typology classifies the ways to visually integrate physical and abstract data into three categories:
    (1) in linked views, the physical space and abstract space are separated; 
    (2) in embedded views, the physical space intersects with the abstract space;
    and (3) in mixed views, the physical and the abstract spaces are placed adjacent so that the
    content can cross from one space to another space.
}
\vspace{-5mm}
\label{fig:views-comparison}
\end{figure*}

\textbf{Linked view} (\autoref{fig:views-comparison} (1)), or coordinated multiple views, 
has been widely used in urban visual analytics
and become standard approaches to display temporal and spatial data~\cite{space-time-visual-anlytics}.
In a linked view, the physical and abstract spaces are separated.
Typically, in a visual analytics system with linked views, the views for physical (generally the maps)
and for abstract data (generally the information charts) are often displayed in parallel.
These views are usually synchronized to display various aspects of the data of an object.
Linked view is very easy to implement and introduce less occlusions.
Although linked view is a powerful and effective visualization method in urban visual analytics, additional
screen real estate is required to show the views. 
Moreover, linked view requires users to mentally relate the physical data to the abstract data,
thereby leading to a considerable mental burden of memory context switching.

\textbf{Embedded view}, or integrated view (\autoref{fig:views-comparison} (2)), 
displays physical and abstract data in the same view,
providing users with a coherent experience when exploring the data. 
In an embedded view, the physical space intersects with the abstract space.
Embedded view has also been widely used in urban visual analytics. 
In most of existing visual analytics systems with embedded views,
the abstract space is encapsulated in the physical space.
Any kinds of spatial data can be visualized using the embedded view,
and these data can be drawn on the map based on locations.
However, the visual representations of data in embedded views usually are basic geometry with colors, 
such as points, lines, flow map, contour map, etc. 
Embedding the complex information charts in a map, which may lead to severe occlusion of other useful
map information, is difficult.
Moreover, unjustified embedded view potentially causes harmful mutual interference between physical and abstract data.

\begin{figure}[tb]
 \centering 
 \includegraphics[width=\columnwidth]{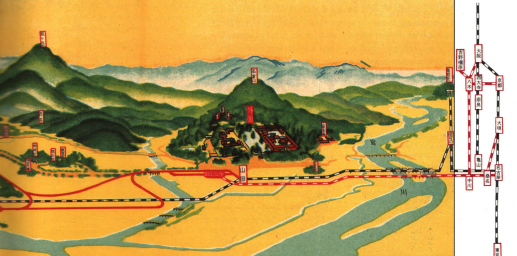}
 \caption{Guide for Vistors to Ise Shrine. In this figure, the railway
    extends from the detailed map view, crosses the boundary,
    and merges into the overview of the national railroad system.
}
\vspace{-5mm}
\label{fig:guideforvistors}
\end{figure}

\textbf{Mixed view} is a novel form to visually merge physical and abstract data.
In a mixed view (\autoref{fig:views-comparison} (3)), the physical and the abstract
spaces are placed adjacent to each other in a proper way for the physical data to move across the boundary
 of physical space and to ``grow'' into the abstract space.
Figure~\ref{fig:guideforvistors} presents a famous example of the mixed view
from the book \emph{Envision Information} by Edward Tufte~\cite[Chapter~1]{Tufte:mixedview}.
In this Japanese travel guide, the railway
extends from the detailed map view, crosses the boundary, and merges into the overview of the national railroad
system.
A person may consider the mixed view as an explicitly linked view. However, in a mixed view, the contents in the
physical and abstract spaces are visually continuous, whereas these contents are visually discrete in a linked view.

We briefly summarize the advantages of the mixed view.
First, compared to the linked view, although the mixed view also separates physical
and abstract spaces, this method allows users to switch context more smoothly and naturally.
Second, the mixed view is more informative than the embedded view because the mixed view avoids the occlusion and mutual
 interference problems, giving designers a free hand to design information visualizations.
Third, by merging physical and abstract data harmoniously, the mixed view engages users with vivid visualizations,
thereby stimulating the interest of users in the data. 
The mixed view is not only a useful tool for data analysts to dig insights,
but also an excellent communication design to present a complicated data to laymen.

However, the mixed view still has certain limitations.
First, the design space of the mixed view is uncertain.
Second, some accessorial techniques are still waiting to be developed. 
For example, we will need a new layout algorithm that ``grows'' physical data from physical space to abstract space. 
A new deformation algorithm that reconfigures the data for a good mixing is also needed.
These limitations are mainly caused by the lack of relevant studies.
We believe these limitations will be tackled in the future with an extensive investigation.

\subsection{Design Considerations}
Based on our typology,
when designing visualizations for physical and abstract data, the designers must select a proper
way of visualizing the hybrid data from the \emph{linked, embedded}, and \emph{mixed views}.
We propose two plain and comprehensive design considerations, namely \emph{visual geometry} and 
\emph{spatial distribution}, that fully utilize the characteristics of data 
to help designers shortlist the candidates.

\textbf{Visual Geometry} is the first design consideration.
The geometry of  	data determines the appearance of the data.
The geometry of physical data (\emph{PG}) is provided, whereas that of abstract data (\emph{AG}) is selected.
For example, the geometry of a road in the map is a line, 
whereas the geometry of the traffic flow data could be a line in a line chart or a bar in a bar chart.

\textbf{Spatial Distribution} of data in its visual space is the second consideration. In this study, we only
consider the relative position of data in spaces, which are not changed with the movement of the space.
We consider this factor because the spatial distribution determines whether the data in two spaces
 are visually coherent or not.
Hereafter, we refer to the distribution of physical and abstract data as \emph{PD} and \emph{AD}.

\begin{table}[]
\centering
\caption{Rules for selecting a view based on our design considerations}
\label{tab:design-considerations}
\begin{tabular}{|c|c|c|}
\hline
         & PD == AD      & PD != AD    \\ \hline
PG == AG & Embedded View & Mixed View  \\ \hline
PG != AG &               & Linked View \\ \hline
\end{tabular}
\vspace{-6mm}
\end{table}

With these two design considerations, we can quickly decide which kind of view we should use based on 
some heuristic rules (\autoref{tab:design-considerations}).
For an object we want to visualize:
\begin{itemize}
\item If the \emph{PG} and \emph{AG} of an object have the same geometric primitive, and its \emph{PD} and \emph{AD} are consistent,
then the embedded view should be used;
\item else if the \emph{PG} and \emph{AG} of an object have the same geometric primitive, but its \emph{PD} and \emph{AD} are inconsistent,
then the mixed view should be used;
\item else if the \emph{PG} and \emph{AG} of an object have a different geometric primitive, and its \emph{PD} and \emph{AD} are inconsistent,
then the linked view should be used.
\end{itemize}

\begin{figure}[tb]
 \centering 
 \includegraphics[width=\columnwidth]{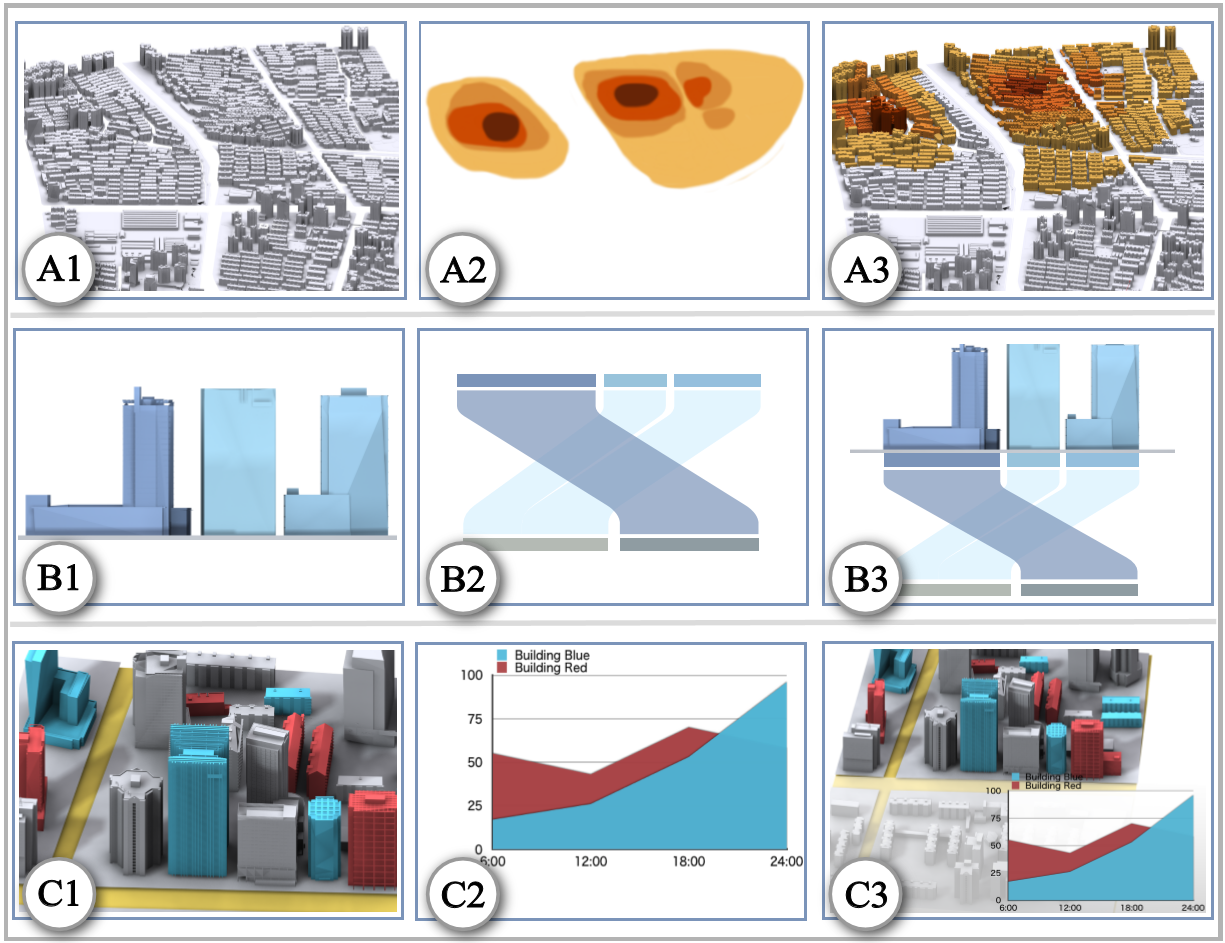}
 \caption{Choosing a way to visually integrate physical and abstract data based on our design considerations, namely, visual geometry and spatial distribution.
 Top: The situation to select the embedded view.
 Middle: the situation to select the mixed view.
 Bottom: the situation to select the linked view.
 }
 \vspace{-6mm}
\label{fig:visualdesign-examples}
\end{figure}

Figure~\ref{fig:visualdesign-examples} shows some examples to effectively illustrate our main idea.
The first row shows the case of using the embedded view to show the data of a region. 
A1 is a map of a certain region, and A2 is a KDE map that displays the population density of the region.
The use of the embedded view (\autoref{fig:visualdesign-examples} (A3)) is suitable because the geometry
of the region and KDE map are both planes, and the distributions of locations in the region and kernels in the KDE map are consistent. 

The second row displays the case of the mixed view, showing the data of buildings.
B1 indicates the three buildings on a 3D map.
B2 is a Sankey diagram that shows the energy consumption of buildings in two months.
Both geometries of these data are planes, but the distributions are inconsistent.
Therefore, we use a mixed view to integrate these two data together. 
The last row presents the case of the linked view to show data of on several buildings. 
C1 displays several coloured buildings divided into two categories.
C2 is a stacked area chart that depicts the statistics of different categories. 
The conclusion that the geometry and distribution of the region in C1 and C2 are different is easy. 
Thus, a linked view is used.

\subsection{Geometry Transformation}
In \autoref{tab:design-considerations}, an empty cell corresponding to the conditions that
PD and AD are consistent, whereas PG and AG are inconsistent. We believe that when the data fall into
this situation, the three views are inappropriate for visualizing the data. For example, the first row
in \autoref{fig:visualdesign-examples} indicates that the KDE map
can be converted to multiple bars distributed on a 2D plane. However, embedding these bars directly 
onto the 2D map, which may cause occlusion and visual clutter problems, is an inappropriate choice.
This situation can be resolved by selecting another visual representation of abstract data to change the
AG or applying certain techniques to deform the PG, such as~\cite{road-bordern-17}, which broaden a road to 
convert its geometry from line to plane.

\section{Research Opportunities}
\label{section:futurework}

Given the immersive analytics is a new research thrust emerging recently,
few research works have been proposed related to immersive urban analytics.
Based on our typology, 
we survey existing works related to immersive urban analytics from academia and industry.
We regard these works as an initial step to identify the research opportunities to be further explored and pursued.
In the following section, we will discuss the technical challenges and future research.

\subsection{Adapt the Linked View}

\begin{figure}[tb]
\centering 
\includegraphics[width=\columnwidth]{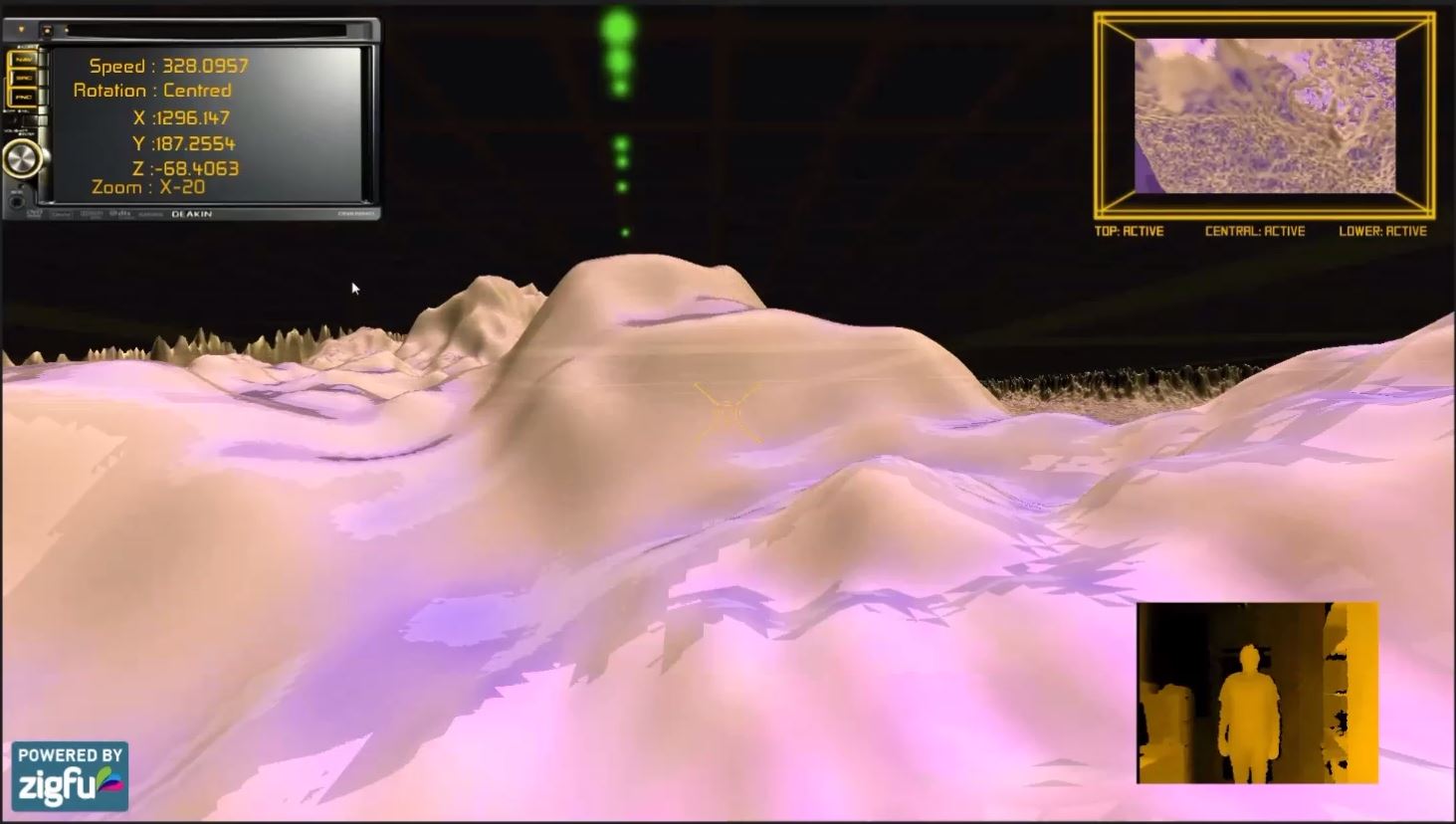}
\caption{Cybulski et al. used linked view in their system~\cite{linked-view:immersive-and-collaborative-analytics} to display
additional digital information besides the physical data.
}
\vspace{-5mm}
\label{fig:linked_view_example}
\end{figure}

Linked view has been widely used in urban visual analytics on traditional Personal Computer (PC).
Many researchers use linked views to display temporal information,
Ferreira et al. built TaxiVis to visualize data on taxi trips including origin-destination (OD) data~\cite{urban-point-taxi}.
In their system, a line-chart is used to show the weekly number of taxi trips in New York City.
Andrienko et al.~\cite{urban-time-graph} visualized multiple trajectories by combing a time graph with a map.
Guo et al.~\cite{urban-guosiming} developed a visual analytics system with a stacked graph and a scatter plot in linked view for temporal data.
In addition, spatial data can be visualized in a linked view. Andrienko et al. proposed a methodology
that converts the trajectories of individuals from geographical space into an abstract space to analyze
the movement behavior of individuals, whose positions are mapped onto the abstract space as points that
move in a group~\cite{urban-point-group-space}.
Crnovrsanin et al. proposed the notion of proximity-based visualization to map the multi-dimensional
data set consisting of proximity values to an abstract space instead of the given spatial layout~\cite{urban-line-proximity}.

Although the linked view becomes one of the standard approaches to display data in urban visual analytics,
it has rarely been used in immersive environments.
Accordingly, one interesting future research direction is to adapt the linked view designs to immersive environments.
Recently, some initial efforts have been done in this direction.
Cybulski et al. presented a project~\cite{linked-view:immersive-and-collaborative-analytics} 
that investigates how immersive environments could be used to support complex collaborative decision-making
through the analysis of intuitive 3D models of business data. 
Their system are developed in VR environments (\autoref{fig:linked_view_example}),
using two linked views to visualize additional abstract data.
However, the linked views in their system are not interactive.
Given the interaction methods in immersive environments are various from traditional PC environments,
it is critical to revise some linked view designs to better suit the new environments.
For example, Urbane~\cite{urban-vis-urbane} adopted a parallel coordinates linked view
in which the users can brush the plot to filter or highlight data by using mouse.
However, it is difficult to brush the plot using gesture,
since the accuracy and sensitivity of gesture cannot be guaranteed.

Another potential research opportunity is to utilize the linked view in other immersive environments, 
e.g., AR and MR.
Different from the VR environments,
the AR and MR environments contain real world objects,
which should be addressed differently from digital content ranging from interaction to visualization.
The interaction and visualization methods should not only provide natural and intuitive user experience,
but also help users to distinguish the digital content from real objects.

\subsection{Improve the Embedded View}

\begin{figure}[tb]
\centering 
\includegraphics[width=\columnwidth]{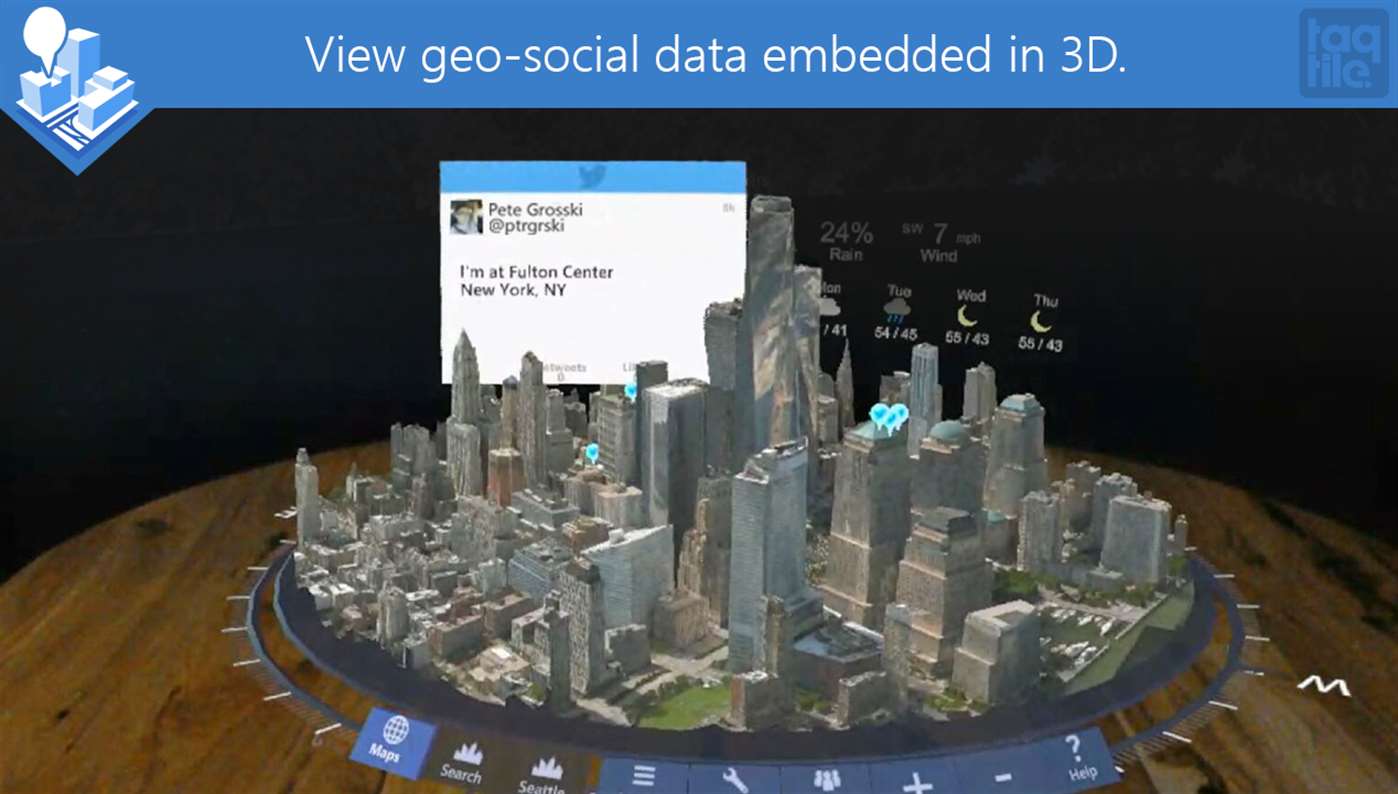}
\caption{The HoloMaps~\cite{Holomaps} can display real-time urban data.
Examples include weather information and tweets with geo-tags on Microsoft Bing Maps 3D 
are embedded in the 3D city model
to allow users to explore a city in an immersive environment.
}
\vspace{-5mm}
\label{fig:holomaps}
\end{figure}

Embedded view is a popular choice for displaying urban data in immersive environments.
Several immersive urban analytics projects adopt the embedded views in their visualizations.
Yoo utilized 3D hexagonal grid cells, 
which are overlaid directly on the 3D map, 
to reveal the geo-tagged tweets related to anti-government protest in London~\cite{embedded-view:yoo2016mapping}.
Moran et al. attempted to study the geographic distributions of Twitter data by developing a 3D 
application utilizing an HMD VR setting~\cite{ia-hmd-twitter}.
However, they only displayed tweets with geo-tags on a 3D map. 
Users can explore those tweets through simple interactions.
Treyer et al. proposed a narrative animated video for urban data analysis.
They created animated visualizations of urban data, such as landmarks and traffic information, 
and then composited it over video footage using Blender~\cite{embedded-view:urban_in_MR}.
Based on Microsoft HoloLens, HoloMaps~\cite{Holomaps} has been introduced to display real-time urban data.
Examples include traffic flow and tweets with geo-tags are visualized 
on Microsoft Bing Maps 3D to allow users to explore a city in an immersive environment.
It still has several possible avenues for future work to improve embedded views.

One possible avenue is to address the occlusion problem caused by intersecting between the physical and abstract space.
Most of the existing methods adopt a strategy that directly overlays abstract data on top of physical data,
which may lead to occlusion problems.
For example, although HoloMaps~\cite{Holomaps} can display rich real-time urban data (\autoref{fig:holomaps}),
the visualizations of abstract urban data introduce annoying occlusions which
block both the physical data and other abstract data.
These occlusions force users to spend more time to navigate the visualization during analysis.
Moreover, in some cases,
the users cannot get rid of the occlusions no matter how he/she change his/her viewpoint.
How to properly tackle the occlusion problems in embedded view is a problem waiting to be solved.

Another possible avenue is to extend the design space of embedded view.
The visual representations of data in embedded views usually are basic geometry with colors, 
such as points, lines, flow map, contour map, etc. 
Embedding the complex information charts in a map is difficult,
since it may lead to severe occlusion of other useful map information
and mutual interference between physical and abstract data.
A promising research direction is to extend the design space of embedded view
by utilizing the power of immersive technologies.


\subsection{Explore the Mixed View}
A study on mixed view would be valuable because existing works related to this field are limited.
The mixed view is rarely used in urban visual analytics. 
To the best of our knowledge, 
we only find a mixed view example from manual design (\autoref{fig:guideforvistors}).
The design space of mixed view is inaccessible. 
The answers to the primary questions,
such as when to use mixed view, how to use mixed view, 
and what are the types of mixed view, are still unclear.
Additionally, the accessorial techniques of mixed view are lacking,
such as layout algorithms to generate mixed view
and interaction techniques to support users to analyze and consume data through mixed view.

\section{Discussion}
\label{section:discussion}
Immersive technologies are attracting serious attention from the field of visualization because of the 
popularity of low-cost-high-quality  immersive modalities, such as HTC VIVE, Oculus Rift, and Microsoft HoloLens.
These revolutionary technologies provide users with more engagements and organic experience; to quote
Mathew Schlepper~\cite{vr:the_state}, ``It's man meets machine, but what happens is strictly within the mind."
Although these modalities themselves are not new, 
researchers' interest is renewed because such low-cost consumer hardware present more opportunities than ever before
to deliver research contributions related to immersive environments to end users.
Given the native support of 3D visualization,
the rise of immersive technologies provides good opportunities to significantly expand the  frontiers of visual urban analytics.

In the wake of this surging wave of immersive technologies, 
we take the first step to explore the possibilities of utilizing these promising technologies in urban science.
Our works blaze a trail in thinking how physical and abstract data can be integrated visually
from the perspectives of visual space instead of detailed visual designs.
The significance of our works is revealed from multiple perspectives.
We first propose an abstract model that characterizes the visualization of immersive urban analytics.
Our model is concise and comprehensive, covering all three types of immersive environments (VR, AR, and MR).
Based on this model, we deduce a typology that sorts out the ways to design hybrid physical and abstract data visualization.
Additionally, our typology helps us recognize the mixed views to visually integrate physical and abstract data
that takes the advantages of both linked views and embedded views. 
As the research field of mixed view is inaccessible, 
this work can hopefully shed new light and arouse interest for further research.
We also propose two plain and comprehensive design considerations
to assist designers in designing visualizations that integrate physical data and abstract data.
Our typology and design considerations can be applied in both immersive environments and traditional PC environments.

However, our work still involves certain limitations.
First, most of the immersive technologies utilize stereoscopy to create stereograms,
presenting two offset images separately to the left and right eyes of a user.
For simplification, our model treats these double image planes as one image plane.
It remains unknown that whether human perceive the stereograms in a different way
from the traditional images in a monitor screen. Our model can be refined along with further research on
the human cognition of stereograms. 
Second, only the conditions of consistency and inconsistency are included in the design considerations.
However, in some sophisticated information visualization, the visual 
geometry of an object can be vary, indicating that the visual geometry in abstract space could be 
partially consistent with the visual geometry in physical space. 
Third, the design considerations only cover the basic situation where only one view needs to be addressed.
The method of designing multiple views in an immersive urban analytics system requires a further study.
Nevertheless, with the rapid development of immersive technologies in the near future, 
these drawbacks can be tackled by further research.


\section{Conclusion}
\label{section:conclusion}

In this work, we study and explore the ways to design hybrid physical and abstract data visualizations
in immersive urban analytics,
which to our knowledge, is the first attempt at systemically characterizing the problem.
An innovative theoretical model is introduced to characterize the visualizations in immersive urban analytics.
A new typology is deduced based on the model, identifying the ways to 
visually integrate physical and abstract data in an immersive environment. 
Following the typology, two design considerations are proposed to guide
designing hybrid physical and abstract data visualizations.
Using our typology and design considerations, designers
can readily select a proper way of integrating the visualizations of physical and abstract data.
Several examples are given to demonstrate the usefulness of our design considerations.

There are several possible avenues for future work. 
A study on mixed view would be valuable because existing research related to this field is limited.
We also plan to use our typology to guide the design of immersive urban analytics applications based on real-world data set.

\newpage

\bibliography{mybibfile}

\end{document}